\documentstyle{mn}

\begin{document}
\title{The structure of spiral galaxies - I. Near-infrared properties of bulges, disks and bars}
\author[M.S. Seigar \& P.A.James]{M.S. Seigar \& P.A.James\\
Astrophysics Group, Liverpool John Moores University, Byrom St., Liverpool,
L3 3AF, U.K.\\
email: {\tt mss} \& {\tt paj} {\tt @staru1.livjm.ac.uk}}
\maketitle

\begin{abstract}

We present data for a sample of 45 spiral galaxies over a range of Hubble
types, imaged in the near--IR J and K bands. Parameters are calculated
describing the bulge, disk and bar K-band light distributions, and we
look for correlations showing the interrelation between these components.

We find that bulge profiles are not well-fitted by the classic de
Vaucouleurs profile, and that exponential or $R^{1/2}$ fits are
preferred. Bulge-to-disk ratio correlates only weakly with Hubble type.
Many of the galaxies show central reddening of their J--K colours, which
we interpret as due to nuclear starbursts or dusty AGN.

We define a new method for measuring the strength of bars, which we call 
Equivalent Angle. We stress that this is better than the traditional 
bar-interbar contrast, as it is not subject to seeing and resolution effects.

Bars are found in 40 of the 45 galaxies, 
9 of which had been previously classified
as unbarred.  Bar strengths are found not to correlate with disk surface
brightness or the presence of near neighbours, but a tendency is found
for the most strongly barred galaxies to lie within a restricted,
intermediate range of bulge--to--disk ratio.  
Bar light profiles are found to be either
flat or exponentially decreasing along their long axes, with profile type not
correlating strongly with Hubble type.  Bar short axis profiles are
significantly asymmetric, with the steeper profile being generally on the
leading edge, assuming trailing arms. In the K--band we find bars 
with higher
axial ratios than have been found previously in optical studies.

\end{abstract}

\begin{keywords} 
galaxies: spiral--galaxies: structure--galaxies: fundamental parameters
\end{keywords}

\section{INTRODUCTION}

\noindent
This paper is the first in a study of the structure of spiral galaxies,
in which we aim to look at the interrelation between bulges, disks, bars
and spiral arms of galaxies over a wide range of Hubble types.  We will
attempt to constrain or distinguish between the various theories 
of formation and evolution of bar
and spiral structures, and our results should also provide useful
determinations of galaxy parameters for input into simulations of the
evolution of galactic structures.  The most controversial question in
this area still concerns the origin of the spiral structure seen in most
gas--rich disk galaxies, which is variously claimed to be due to
quasi--stationary density wave modes (Lin \& Shu 1964, 1966; Bertin et al.
1989a,b), stochastic self--propagating star 
formation (Gerola \& Seiden 1978) or
driving by nearby neighbours or central bars (Kormendy \& Norman 1979),
and we will discuss this in a future paper (Seigar \& James 1997; paper II).
However, there are also unresolved questions concerning the origin of
bars, and of the Hubble sequence of galaxy types, which we
address in the present paper.

Our main technique is to image a large sample of galaxies in the near-IR
J and K bands (centred on 1.3 and 2.2 $\mu$m respectively), which has
become generally accepted as the most powerful technique for revealing
the structure of the dominant stellar mass component in galaxies. The
advantages of this wavelength range are the relative freedom from
extinction effects compared with visual imaging, and that the near-IR
light from normal galaxies arises from the old stellar population which
dominates the stellar mass. Whilst this latter claim has sometimes been
contested on the grounds that luminous giants and supergiants could in
principle dominate the near-IR light of a galaxy with recent star
formation, Rix \& Rieke (1993) indicate that the
supergiant contribution to the K-band light from normal galaxies is at
most 15--25\%, a small effect compared to the impact of extinction and
star formation on the visual morphology of star--forming galaxies.

Quantitative measurements will be presented for all of the stellar
components of the galaxies (disks, bulges, bars and arms) to enable us to
study the interrelations between them.  This will provide the perfect
database for a test of secular models of galaxy evolution (Combes et al.
1990; Courteau, de Jong \& Broeils 1996) where bar--induced star--formation
and mass transfer provide the central mechanism for the transformation of
disk galaxies along the Hubble sequence. Another aim of this study is to
produce realistic parameters which can be used both as input parameters
for models of disk galaxies, and to compare with the output from such
models.

This paper will describe the acquisition and reduction of the near--IR
imaging data, and the analysis of the properties of the disks, bulges and
bars of the full sample of 45 galaxies. The interrelation between the
properties of these components will be investigated in detail, but spiral
arm structure will be dealt with in a future paper (Seigar \& James
1997, paper II). Section 2 describes the sample selection and observations; 
section 3 describes data reduction techniques;
section 4 describes properties of bulges; section 5 describes properties of
disks; section 6 describes properties of bars and section 7 contains a
discussion and conclusions.

\section{OBSERVATIONS}

\noindent
Observations are presented for a sample of 45 spiral galaxies, selected
using the NASA/IPAC Extragalactic Database\footnote[1]{The NASA/IPAC 
Extragalactic Database (NED) is operated by the Jet 
Propulsion Laboratory, California Institute of Technology, under contract
with the National Aeronautics and Space Administration.}
(henceforth NED) on the basis
of their diameters ($<$1.5 arcmin at the $B_{25}$ isophote) and their
face-on orientations (axial ratio (b/a) $>$0.5, typically 0.8).  They
were specifically selected to span the full range of Hubble types Sa--Sd,
on the basis of their classification in the Third Reference Catalog of
Bright Galaxies (de Vaucouleurs et al. 1991; henceforth RC3), and no
attempt was made to select a statistically representative or complete
sample. A Hubble constant of 50 kms$^{-1}$Mpc$^{-1}$ is assumed
throughout.

The data were obtained with the 3.8 metre United Kingdom Infrared
Telescope (UKIRT) on Mauna Kea, using the near-IR camera IRCAM3.  The
observing dates were 1995 February 4--6 and 1995 November 17--19, with
further images being taken on the nights 1996 May 21 and 22 as a backup
for a programme which could not be carried out through the prevailing
thin cirrus.  Conditions were generally photometric for the first two
runs, and the seeing was typically sub--arcsecond.  However, the stability
and guiding of UKIRT at that time were such that images were sometimes
enlarged at the 1 arcsec level by telescope motions, particularly in the
RA direction. This can have the effect of smearing the point spread function
(PSF), but as it is of the order of the seeing, its effect is small in these
observations. IRCAM3 is a 256$\times$256 InSb array camera, with a pixel
scale of 0.286 arcsec, giving an image scale of 73$\times$73 arcsec.  The
galaxies were imaged in the standard J (1.3 $\mu$m) and K (2.2 $\mu$m)
filters.  Typical total exposures were 9 and 30 minutes per galaxy at J
and K respectively, with each exposure being split into several
"jittered" sub-exposures to facilitate bad pixel removal. These 
observations result in a detection of 20.5 magnitudes per square arcsec at
1$\sigma$ per pixel at K
and 21.5 magnitudes per square arcsec at J. Equal times
were spent observing nearby blank sky in the same jitter pattern to
produce a median sky flat for each galaxy.  Near-IR standard stars from
the UKIRT faint standards list were observed throughout the photometric
nights for calibration purposes. Photometric  errors are calculated  to  be
 at the 3\% level. Photometry was also checked against previous work. Only one
galaxy (NGC 6574) had been previously observed 
in these wavebands before and this had
been using aperture photometry and has been reported by Balzano \& Weedman 
(1981). They report values within a 10.1 arcsec aperture of a K-band 
magnitude of 10.14\underline{+}0.10 and a J-band magnitude of 
11.28\underline{+}0.05. Within the same aperture we get a magnitude of
10.13\underline{+}0.04 at K and a magnitude of 11.24\underline{+}0.16 at J.

\section{DATA REDUCTION}

\noindent
Initial data reduction was entirely standard and will therefore be
described only briefly.  The STARLINK packages KAPPA and FIGARO were used
for basic image processing.  All frames were dark-subtracted, with bias
subtraction not being needed due to the non-destructive read mode of
operation of IRCAM3.  Sky flats were constructed by taking medians of
sky frames taken within $\sim$30 minutes of the given
galaxy observations, since experimentation showed that variations in
either the array or in the sky colour gave increased noise if more sky
frames were included in the median.  The individual galaxy integrations
were then flat-fielded, and all known bad pixels set to "magic" values
such that they were ignored when the images were added to give the final
mosaics.  The individual images were sky-subtracted prior to this final
addition, by taking a mean of the median brightnesses over a 25x25 pixel box
in each of the four corners of all of the frames.
The images of the standards were similarly dark-subtracted,
flat-fielded and had their bad pixels removed, prior to photometry being
measured using 12 arcsec software apertures.

The RGASP package was then used to fit ellipses to the isophotes in the
reduced galaxy images. The ellipticity of the outermost ellipses was then
adopted and the PROF routine, which takes medians around ellipses at increasing
radii, was run keeping this ellipticity and position angle constant. 
The REBUILD facility in RGASP enables a simulated
galaxy to be made up using the ellipticity and position angle and the
median surface brightness calculated around these ellipses, and
subtracting this rebuilt image from the original image removes the
elliptically-symmetric bulge and disk components.  This procedure leaves
an image of the bar and spiral arms.  Plotting the surface brightness of
the fitted ellipses as a function of semi-major axis gives the light
profiles shown in figure 1, which will be used to do a one-dimensional
bulge-disk decomposition below.

The bulge-- and disk--subtracted galaxy images were then deprojected in an
attempt to overcome inclination effects on the spiral arm pattern and bar
morphology, using a similar technique to that of Elmegreen \& Elmegreen
(1984).  This was done by measuring the disk major and minor axes at an
isophote well out in the disk, where any ellipticity is plausibly due to
inclination effects (although Rix \& Zaritsky (1995) have demonstrated
that there is measurable ellipticity even in face-on spiral galaxies,
which introduces some uncertainties into this process).  The images were
then rotated to make the measured major axis vertical, and the minor axis
was then stretched by a factor equal to the measured major/minor axis
ratio.  This correction was always quite small, given the generally
face-on nature of galaxies in this sample.

\section{PROPERTIES OF BULGES}

\noindent
Bulge--to--disk ratio (B/D) has always been taken as a key parameter in
defining the position of a galaxy on the Hubble sequence. This ratio can
be significantly affected by extinction in the visual, and K--band imaging
clearly gives a much more reliable indication of the stellar mass in the
disks and bulges of spiral galaxies.  B/D was calculated for all the
galaxies in the present sample, using a one-dimensional profile deconvolution 
technique on the K-band light profiles (figure 1).  De
Jong (1996a) has demonstrated that there are some advantages in doing a
full two--dimensional fit to the galaxy images when determining disk and
bulge profiles, but the advantages are not huge and outweighed, for the
present study, by the greatly increased complexity of the fitting
process.

In order to separate the bulge and disk components of a measured light
profile, some assumptions have to be made about the functional form of
the individual components, and the reliability of the results necessarily
depends on how good an approximation these forms are to the true light
distributions (see Capaccioli \& Caon 1992 for a general discussion of
profile decomposition techniques).  Traditionally disks have been
represented by an exponential light distribution and bulges by the
$R^{1/4}$ light distribution found by de Vaucouleurs (1953) to be an
excellent fit to the light profiles of most elliptical galaxies.
Recently, however, Andredakis \& Sanders (1994) and de Jong (1996b) have
demonstrated that most bulges, particularly those of late-type spirals,
are better represented by exponentials, with a scale length about an
order of magnitude smaller than that of their disks. This follows earlier
suggestions that NGC~4565 (Frankston \& Schild 1976) and the Milky Way
(Kent, Dame \& Fazio 1991) have exponential bulges.  Even the bulges of
early-type galaxies were found by de Jong (1996b) to be better fitted by
either exponentials or $R^{1/2}$ profiles, where the latter are
intermediate between the exponential form and the much cuspier 'de
Vaucouleurs' profile.  For our study we elected to leave the index in the
fitted bulge light profile as a free parameter, which had the dual
advantage of allowing the best fit to the bulge profile to be found, and
of enabling us to investigate any possible change in the best-fit bulge
index with B/D ratio, as indicated by de Jong (1996b). The fit was performed
using a Levenberg-Marquardt method which reduces the value of $\chi^{2}$ of
a fit of the analytical profile to the measured surface brightness, $\mu$, as
a function of radius, $r$. The method
fits to data and outputs the bulge half-light radius and surface brightness at 
that radius, disk central surface brightness, disk scale length, disk 
central  surface brightness and bulge-profile index for which it finds the 
lowest value for $\chi^{2}$.

An important issue in galaxy profile fitting is atmospheric seeing, which
is particularly likely to affect the measured bulge parameters by
smoothing out the light from the central, fairly cuspy region.  We
quantified the effects of seeing on measured parameters by constructing
artificial galaxies with profiles typical of the galaxies in our sample,
smoothing by a Gaussian to simulate seeing effects, and then adding noise.  
These artificial galaxies were then analysed using the same
data reduction process as for the real images, to assess how well the
software can extract the profile parameters, both with and without the
effects of seeing.  The results are given for two of these simulated
galaxies in Table 1, where the values given are the input and recovered
values of bulge surface brightness, half-light radius and power law, disk
central surface brightness and disk scale length, with simulated seeing
of 0.85 arcsec, typical of the real images. (With no simulated seeing,
all parameters were recovered without error.) Simulated galaxy (a) has
bulge parameters typical of the galaxy sample, whereas (b) represents
a worst case, with a low-luminosity, short scale-length bulge (B/D$\sim$
0.08). Table 1 shows that the effects of seeing are small, affecting
measured parameters by at most 10\%, which is likely to be within the
random errors. Effects on disk parameters were found to be of a similar
order. Thus no specific seeing corrections were applied to these images,
or to any parameters derived from them. This technique also allowed us to 
estimate errors for the fitted parameters. However, where the disk starts
to get down to the sky level, errors in determining the sky level have been 
assumed to be the
dominant source of error in determining the disk parameters. As a result of
these systematic effects we believe that the typical error associated with our
measurement of the bulge-profile index is about $\underline{+}$0.05.

The measured parameters of the bulges are shown in table 2. These have been 
converted into physical quantities using distances calculated assuming a 
constant Hubble flow. Note that the profile fit of the galaxy ESO 555 --G 013
revealed that this was almost completely dominated by disk and as a result the
fit could not produce a stable fit to the bulge. We could therefore not 
constrain the bulge parameters for ESO 555 -G 013 and thus a bulge-to-disk 
ratio was not calculated.

Looking at the derived properties of the bulges, we first confirm
the conclusion of de Jong (1996b) that better bulge fits
are given by a bulge-profile index between 1/2 and 1 (the two values adopted 
by de Jong) than with the de Vaucouleurs R$^{1/4}$ profile.
Indeed, there is
some evidence for a bimodal distribution, with indices around these two
values being preferred. However, no strong correlation was found 
(correlation coefficient=0.33; significance=97\%) between
this index and Hubble type (figure 2), whereas de Jong found that values
less than 1 tended to be found only in early-type galaxies. This distribution
is in good agreement with that found by Andredakis, Peletier \& Balcells (1995)
over the same range in Hubble type. There is
also no significant correlation between best-fit bulge index and bulge
luminosity (correlation  coefficient=0.17; significance=76\% - see 
figure 3), which differs from the case for dwarf elliptical
galaxies, where smaller power-law indices and cuspier profiles tend to be
found in the more luminous galaxies (Young \& Currie 1994; James
1994).

\subsection{AGN or starburst contributions to bulge light}

A final concern with the bulge profiles is that they may be contaminated
by emission from active galactic nuclei (AGN). It is hard to constrain
such a component simply from the light profile, since even quite a
prominent point source can be indistinguishable from a cuspy bulge light
distribution.  However, colours do provide a useful diagnostic, since AGN
are generally heavily extinguished, giving much redder near-IR colours
than the surrounding stellar bulge (Glass \& Moorwood 1985).  We thus
combined central J and K light profiles to give radial colour maps for
all galaxies, which we could use to search for any trend towards red J--K
colours in the very central regions. The seeing for each frame in each band was
calculated using the PSF routine in the STARLINK KAPPA package. It was 
typically found to be approximately the same in both the J- and K-bands, at
$\sim$0.8 arcsec. The typical photometric errors in 
the central colour changes are therefore calculated to be at the 1\% level. 

Representative J--K profiles are shown in figure 4.  Disk colours
generally lie between J--K of 0.7 and 1.0, consistent with the
expectations for a fairly blue, star-forming system.  Approximately 50\%
of our galaxies show no central colour change whatever at the 0.1 mag
level (e.g. UGC~3296 in Figure 4).  Of the remainder, the nuclear colour
changes vary between a just-detectable increase of $\sim$0.1 mag to over
a magnitude, with the reddest nuclear colours approaching J--K$=$2.0
(UGC~3707 and IC~568, shown in figure 4). We never see a nucleus bluer
than the surrounding disk.

J--K values between 1.0 and 2.0 could be the result of either a dusty
starburst, or of a dust-embedded AGN.  Glass \& Moorwood (1985) show
that JHKL colours have some power in distinguishing between these
possibilities, but we do not have the required observations to apply this
to our galaxies.  However, there does appear to be useful information in
the radial profile of the nuclear colour change.  Comparing the colour
profiles of IC~568 and UGC~3707, it is clear that the redward trend
starts much further out for IC~568, whereas the red nucleus of UGC~3707
is very small, and almost certainly unresolved.  Thus we claim that the
latter galaxy probably hosts an AGN, and the former an extended starburst
nucleus.  Supporting evidence for a nuclear starburst in IC~568 comes
from its having a very strong bar, a feature often linked with enhanced
central star formation (e.g. Roberts, Huntley \& van Albada 1979; Huang
et al. 1996). (However, it should also be noted that UGC~3707 has a very
strong bar.)  We also show in Paper II that IC~568 has the second largest
far-IR luminosity of any of the galaxies studied here, which is almost
certainly linked to its nuclear activity.

Of the galaxies showing clear nuclear reddening ($\Delta (J-K)>$0.2), 7
show evidence for an extended red component (IC~357, NGC~2503, IC~742,
IC~1809, IC~1196, UGC~3806 and IC~568, in order of increasing strength),
and 3 have unresolved red nuclei (UGC~850, UGC~3839 and UGC~3707)

\section{PROPERTIES OF DISKS}

\noindent
The bulge/disk deconvolution described in the previous section also
yielded parameters for the best-fitting exponential to the disks of all
galaxies in the sample.  Given that we chose to use only exponentials,
there are only two parameters in the disk fit for each galaxy, the
central surface brightness and the scale-length in kpc of the best-fitting
exponential.  These are tabulated in columns 2 and 3 of Table 3, which
summarises the disk and bar parameters for the 45 spiral galaxies.
Columns 4 and 5 give the disk K-band absolute magnitude and K-band
bulge-to-disk ratio respectively.

The disk central surface brightnesses have been corrected using the following
equation given by de Jong (1996),
\begin{equation}
\mu^{i}=\mu-2.5Clog\left(\frac{a}{b}\right)
\end{equation}
where $\mu^{i}$ is the corrected surface brightness, $\mu$ is the measured 
surface brightness, $a$ is the major axis, $b$ is the minor axis and $C$ is
a factor dependent on whether the galaxy is optically thick or thin. If
$C=1$ then the galaxy is optically thin, if $C=0$ then the galaxy is optically
thick. De Jong (1996b) uses the optically thin case and we adopt this approach.

The distribution of disk central surface brightnesses has been analysed
many times since the classic study by Freeman (1970), who discovered a
very narrow measured range in this parameter, since termed the Freeman
Law.  This has recently been comprehensively reanalysed by de Jong
(1996b), who used near-IR photometry to minimise possible extinction
effects.  Whilst de Jong confirmed that there is a sharp cutoff in the
high surface brightness side of the distribution, he found that after
accounting for selection effects, there was no such cutoff on the low
surface brightness side.  Thus the overall distribution in disk surface
brightness is much broader than claimed by Freeman (1970).

The present sample of galaxies was not selected as an unbiased
statistical sample for the type of study undertaken by de Jong (1996b).
However, it is still instructive to compare the distribution of disk
parameters with those found by de Jong, prior to his application of the
selection effect corrections.  Figure 5 shows disk central surface
brightnesses plotted against total disk luminosity, which can be compared
with the distributions in both parameters found by de Jong.  We find a
range of disk K absolute magnitudes predominantly lying between --22 and
--26, with two fainter disks at M$_{K}\sim$--21. De Jong (1996b) finds
exactly the same range and distribution in disk luminosity, once the
different assumed Hubble constants are taken into account.  The agreement
for disk central surface brightnesses in the K-band is somewhat less
good, however.  We find a range of just over 3 mag. in this parameter,
16.1--19.5~mag., and also find absolutely no correlation with Hubble type
(figure 6).  Comparing this distribution with figure 3 of de Jong
(1996b), he finds a wider range, extending to both higher and lower
surface brightnesses.  Almost all of his very low surface brightness
disks are extreme late--type galaxies which we do not have in our sample,
but the excess of high--surface--brightness disks, at $\sim$16 K
mag./arcsec$^2$, is harder to understand.  All of these galaxies in de
Jong's sample lie between Hubble types Sb--Sc, and figure 6 shows that
we detect no such galaxies.  However, on the whole there is no evidence
for any systematic offsets in photometry between our data and de Jong
(1996b), and any differences are probably due to sample selection.

In Hubble's classification scheme, the position of a galaxy along the
sequence was thought to be determined largely by the relative dominance
of the bulge and disk.  However, figure 7 shows that this correlation is
fairly weak when near--IR parameters are used.  This figure shows the
fraction of the total K-- band luminosity of the galaxy which is
represented by the disk, plotted as a function of catalogued Hubble type
from RC3.  In general, the later type
galaxies do appear somewhat more disk-dominated, as Hubble's scheme
would predict, but there is a large scatter in this correlation, and
again we confirm a conclusion of de Jong (1996b), in that B/D ratio is
not in fact the central parameter in determining Hubble type.  De Jong
(1996b) concludes that spiral arm type must instead be the main
determinant of morphological type, a suggestion which we will discuss in
Paper II.

In figure 8 we show that there is a weak correlation between
disk scale lengths and bulge half--light radii (correlation coefficient
0.50, significance 99.83\%) with a mean ratio of 5.4.  
This is a very similar result to that found
by Courteau et al. (1996), although they chose to plot the measured
scalelengths in arcsec to avoid bias by resolution effects.  We feel that
biases caused by the spread of distances to our galaxies will be more
significant in potentially causing spurious correlations, so we prefer to
plot absolute scalelengths.  Another difference is that Courteau et al.
fitted bulges by exponentials only, whereas we leave bulge index as a
free parameter, but again this change does not destroy the correlation.
Courteau et al. (1996) interpret the correlation as evidence for secular
models, where bulges are assembled gradually through a series of bar
instabilities, as this theory directly links properties of disks and
bulges.  It is also interesting to note that Aaronson, Huchra \& Mould
(1979), in one of the first attempts to find a physical basis for the
Tully-Fisher relation (Tully \& Fisher 1977) between luminosity and
rotational velocity in spiral galaxies, found it necessary to invoke a
universal mass profile for spirals, which is not a strong function of
Hubble type. Figures 7 and 8 may provide some evidence for this universal
profile in the K-band light distributions.

Figure 9 shows the distribution of disk central surface brightness as a
function of disk exponential scale length, which shows a reasonably
significant correlation (correlation coefficient 0.56, significance
99.97\%). Again, this is very similar to the distribution given in de
Jong (1996b), once differences in assumed Hubble constant are accounted
for.  We confirm the lack of high-surface brightness, long-scale-length
disks, which cannot be the result of selection effects, whereas the
cut-off at low surface brightness certainly is due to selection.

\section{PROPERTIES OF BARS}

\noindent
We now consider the properties of the bar components as revealed by our
imaging, again focussing principally on the K band images to minimise the
effects of dust obscuration.  For simplicity, we adopt as a working
definition of a bar, that part of the central light distribution which is
left when the elliptically-symmetric disk and bulge components have been
subtracted (in this case, using the RGASP REBUILD facility).  Thus we
include in this definition all oval distortions of the bulge and central
disk as discussed by various authors. Also,
there is potential confusion between the outer bar and the inner parts of
the spiral arms or other disk asymmetries, which are also left after the
REBUILD subtraction.  In general, there is a sharp bend or a
discontinuity between the end of the bar and the beginning of the arms
which averts this problem, but there are some cases (e.g. IC~357) with
S-shaped bar/arm morphologies, where the arms sweep back gradually from
the ends of the bar, and the definition of the bar end is somewhat
arbitrary.

The first and very striking result is how common bars are revealed to be
in this galaxy sample.  Of the 45 galaxies studied, 31 are classified as
barred on the basis of their optical morphologies in RC3, a similar
barred fraction to previous optical studies. However, a visual analysis of the
K-band images reveals evidence for bars in a further 9, giving a final
barred fraction of about 90\%.  Our initial selection was not biased
towards barred galaxies, so this fraction should be representative of the
general population of bright spiral galaxies. We note that the
combination of near-IR imaging and the subtraction of the disk and bulge
makes this study particularly sensitive to the detection of weak bars,
which accounts for the higher barred fraction than any previous study of
which we are aware, although there has been a trend for estimates of the
barred fraction to increase with time. De Vaucouleurs (1963) claimed that
just one-third of disk galaxies have bars, whereas more recent estimates
generally put the figure at about two-thirds. Given our present findings,
it is tempting to speculate that bars may always be present in disk
galaxies at some level.

We now attempt to quantify the strength of the bars detected in these
galaxies.  Since there is no generally accepted method for doing this,
and since it is central to our analysis, we have defined our own
procedure for producing repeatable and objective measures of the bar
strength.  In Paper II we will apply essentially the same method to
determining arm strengths in the same galaxies.  The key problem is to
find a measure which is not systematically affected by the quality of the
observational material used, and in particular by the noise level and
seeing effects. This means that, for example, the traditional method of 
taking the peak intensity
at the centre of the bar and ratioing this by the disk intensity at the
same radial distance to produce a bar-interbar contrast is not
satisfactory, since poor seeing conditions will smooth out the bar
emission and decrease this contrast.  The most robust measure we can
think of is closely analogous to the quantity Equivalent Width which is
widely used in spectroscopy, and is specifically designed to be
independent of noise and resolution effects.  We term our analogous
quantity 'Equivalent Angle', henceforth EA, and define it to be the angle
subtended at the center of the galaxy by a sector of the underlying disk
and bulge which emits as much light as does the bar component, within the
same radial limits (see figure 10 for a diagrammatic representation of EA).  
Thus a bar which emitted as much light as the
underlying disk between, say, 0.5 and 1.0~kpc radius would have an EA of
180$^{\circ}$ (not 360$^{\circ}$ since we define EA to refer to one end
of the bar, and the other end in this case would also contribute
180$^{\circ}$) within these radial limits.  We derive EA as a function of
radius, but also take a mean value over the whole detected radial extent
of the bar to parametrise the strength of the bar as a whole, normalised
to the surface brightness of the underlying galaxy. The errors in EA are 
derived from errors in the calculated position angle (PA), ellipticity and sky 
levels. The errors in PA and ellipticity are important because EA is always
calculated in a polar coordinate frame (ln$R$, $\theta$). Both the difference
and rebuild images are converted to polar coordinates. The rebuild image in the
polar coordinate frame should not have variations in surface brightness with
azimuthal angle, but errors in the PA and ellipticity can introduce variations
in this. 

Overall bar strengths,
bar lengths and bar widths are tabulated in table 3.
Figure 11 shows a histogram of the overall EA values for the 40 galaxies
with detected bars, with the shaded areas representing  bar
strengths for those galaxies with nearby neighbours. Neighbours are
defined, somewhat arbitrarily, as galaxies of comparable luminosity to
the observed galaxy, at a projected distance within 6 diameters of the
observed galaxy.  In general redshifts are not available and so chance
line-of-sight associations cannot be excluded.  Probably unsurprisingly,
no strong dependence of bar properties on the presence of such neighbours
is found, supporting the conventional view that bars are intrinsic
features of disk galaxies, and occur as a result of disk instabilities.
The overall bar strengths can be compared with the values quoted by Ohta,
Hamabe \& Wakamatsu (1990), who find that bars comprise between 24\% and
47\% of the total galaxy light within the radii occupied by the bars.
These convert to EA of $\sim$40$^{\circ}$--85$^{\circ}$, which is
consistent with the range of values we find for strongly barred galaxies
(Ohta et al. (1990) specifically selected galaxies with strong bars).

It is then of interest to test whether there are any observable
properties of disks which make them particularly susceptible to the
formation of bars.  Theoretical analysis and simulations suggest that bar
instabilities arise as a natural consequence of cold, self-gravitating
disks, and that the bar mode can only be suppressed by the presence of a
dynamically warm stabilising component, such as a massive bulge or
spheroidal halo (Hohl 1971; Ostriker \& Peebles 1973).  We can test such
effects by correlating bar strength with the
measured properties of the bulges and disks of our galaxies.  Figure 12
shows bar strength plotted as a function of the disk central brightness,
and we find no overall dependence in the expected sense.  Indeed, if
anything the strongest bars tend to be found in the lowest surface
brightness disks, although this is not a significant correlation.  Figure
13 shows that there is also no overall trend between bar strength and
bulge- to-disk ratio, although there is evidence that the most strongly
barred galaxies have B/D values falling in a very narrow range.  The 10
most strongly barred galaxies all have B/D between 0.28 and 0.52,
compared with a full range of B/D from 0.04 to 1.56. A statistical test 
showed that the spread in B/D values for the strongly barred (EA$>30^{\circ}$)
galaxies is narrower than that of the overall sample at the 99.92\% 
($>3\sigma$) level. This is quite a
striking result, and it is hard to see how it would come about as a
result of systematic effects.  Whilst the sample is in no sense
statistically complete, the strength of the bar was not included in the
selection criteria, which were purely defined in terms of Hubble type,
diameter and inclination.  Even if there were some selection systematic
which correlated with classified Hubble type, we have shown that this
latter correlates only weakly with B/D.  The correlation cannot be a
simple effect of the type that bright bars live in bright galaxies
because the definition of bar strength in terms of EA effectively
normalises bar strength to the luminosity of the surrounding bulge and
disk.

It is hard to find any theoretical explanation for strong bars
preferentially inhabiting galaxies of intermediate B/D.  It is still
generally thought that the bar instability results from a cold disk whose
kinematics are dominated by rotation (Ostriker \& Peebles 1973).  The
much discussed secular theory of galaxy evolution (Combes et al. 1990;
Pfenniger \& Norman 1990) postulates that this instability may provide a
mechanism for enhancing the bulge, since bars are known to funnel
material from the disk into the central regions. Accretion of a large
amount of mass in the centre ultimately destroys the bar (Friedli \&
Benz 1993), stopping or at least interrupting the evolutionary process,
but overall this appears to be a viable mechanism for the evolution of
isolated galaxies along the Hubble sequence.  However, this does not lead
to any obvious correlation between bar strength and B/D, except that one
might see a generally declining bar strength with increasing dominance of
the bulge.  Indeed, if secular evolution is the dominant process in
determining the Hubble sequence, then Figure 13 would imply that galaxies
of intermediate B/D should be quite rare, as the strong bars would
rapidly evolve galaxies through this state.  Clearly this simplistic
analysis does not fit the observed distribution of B/D values. 
Another similar plot (not shown) of bar EA against 
the ratio of the disk scale length to
the bulge half-light radius ($h/r_{e}$) also revealed that the strongest 
bars fall in a narrow range of $h/r_{e}$ from $\simeq$4.6 to 6.2. 

We now move on from a discussion of the overall strength of bars to the
light distribution within individual bars.  There has been relatively
little previous work on this subject, and one of the few analytic
distributions proposed for bar light is that of Freeman (1966):
\begin{equation}
\Sigma_{bar}(x,y) = \Sigma_{0,bar} (1-(x/a_{bar})^2-(y/b_{bar})^2)^{\frac{1}{2}},
\end{equation}
where $\Sigma_{0,bar}$ is the bar central surface brightness, and
$a_{bar}$ and $b_{bar}$ are the semi-major and semi-minor axes
respectively.  This gives a profile that is flat in the central regions,
then falling increasingly steeply to zero at $x=a_{bar}$ along the major
axis and $y=b_{bar}$ along the minor axis. Beyond these limits it takes
imaginary values, and must be set to zero.  Observational studies of bar
profiles (Elmegreen \& Elmegreen 1985; Baumgart \& Peterson 1986; Ohta et
al. 1990; Elmegreen et al. 1996) find a range of profile types.  All of
these studies claim that at least some bars show exponential profiles
along their major axes, and the N-body simulations of Miller \& Smith
(1979) provide some theoretical justification for the existence of such
bar profiles.  However, some galaxies are also found to have flat major
axis profiles, which more nearly resemble the functional form proposed by
Freeman (1966). Elmegreen et al. (1996) conclude that the flat bars tend
to be stronger compared to the galaxy luminosity, and are found in
earlier Hubble types, whereas the weaker exponential bars are found
preferentially in late-type spirals.

We measured bar profiles in three ways, starting in two cases from the
bulge- and disk-subtracted, inclination-corrected K-band images.  From
these we calculate both the bar EA (defined above) as a function of
radius, which gives a strength normalised to bulge and disk surface
brightness, and also major and minor axis profiles obtained by summing
flux in slices oriented perpendicular to the appropriate axis.
Finally, we use the non--subtracted K--band images to look at
major axis profiles, so as to compare them with the bar light--profiles
of Elmegreen et al. (1996).

Considering these major and minor axis profiles first,
we confirm some of the conclusions reached by Elmegreen et al. (1996).  
Figure 14 shows three representative K--band bar profiles in both the
subtracted cases (right column) and non--subtracted cases (left column). 
The non--subtracted bar profiles illustrate that we find two
main types of bar profiles. We find flat profiles (IC 357 and IC 568), 
which are characterised by a flat plateau,
with a break at the end of the major axis and a steep descent at
large radii, and weaker, exponentially decreasing bars 
(e.g. NGC 5737). Both types were also found by Elmegreen et al. (1996). 
Also plotted in figure 14 are bar major--axis profiles found for galaxies
after subtraction of the ellipse--fit model (right column). In the case of IC
568, this is seen to rise at low radii, reach a plateau at intermediate radii,
and then decrease at high radii. The bar in IC 568 is classified as flat and
therefore it is not surprising that the bulge-- and disk--subtracted 
bar profile
reaches a plateau. Other galaxies with flat bars have also been found to
have similar bar-profiles after subtraction of the bulge and disk. However,
generally these bar profiles do not reach a plateau, but turn over and start
to decrease immediately (e.g. IC 357).

We do not find a clear-cut correlation between bar profile type and Hubble
type, in disagreement with Elmegreen et al. (1996).  
For the 24 most strongly--barred galaxies we can fairly
unambiguously classify the bars as being either flat or
exponential.  Within this subset, we find that 13
galaxies are of type Sa--Sb, and of these 8 (62\%) have flat bars and 5
(38\%) exponential bars. For the 11 galaxies of types Sbc--Sd, 
5 (45\%) have flat
bars and 6 (55\%) exponential bars.  
Thus no correlation is found between bar type
and Hubble type, although we are clearly suffering from small number
statistics. There is a hint of a correlation between bar type and
B/D ratio in the sense predicted by Elmegreen et al. (1996), since the mean
B/D for galaxies with flat bars is 0.39$\pm$0.16, c.f. 0.21$\pm$0.14 for
galaxies with exponential bars.

The other measure of the bar major axis profiles comes from the radial
dependence of EA, shown for a representative galaxy in Figure 15.
Normalised to the underlying disk and bulge in this way, the bar profiles
generally show a smooth increase in the central regions, where the
dominance of the bulge is decreasing with radius, and a decrease at large
radii where the fall-off is clearly much steeper than the exponential
profile of the disk.  None of the profiles show an extended plateau in
these EA plots as might be expected, for example, were the bar to
represent simply an enhancement of the underlying disk density at a given
radius.  Thus it appears that bars, whether of the flat or LD forms, are
separate entities from the bulge and disk, and with light distributions
which do not relate in any simple fashion to these other components.

Figure 16 shows bar short--axis profiles for some of the more strongly--barred
galaxies in the sample.  These were made by collapsing all of the flux
along the long axis of the bar from the centre out to one end, thus
giving an intensity-weighted mean cross-section of the bar.  These K-band
cross sections are generally smooth but significantly asymmetric, in
marked distinction to the early-type galaxy bars investigated by Ohta et
al. (1990), who found bar cross sections to be highly symmetric.  The
sense of asymmetry for our bars is generally that the leading edge has a
steeper profile than the trailing edge, where these terms are defined
assuming that the bars rotate in the same sense as the spiral arms, and
that the arms are trailing. Figure 16 shows bar cross--sections for the
galaxies IC 1764, IC 357 (typical cases with steeper leading edges), IC 568
(with a steeper trailing adge) and NGC 2529 (approximately symmetric).

We find a very wide range in axial ratios for the bars in our sample,
ranging from $(b/a) = $0.094 to 0.53, where the dimensions used are FWHM
values from the bar (bulge and disk subtracted)
profiles described above, as shown in Figures 14 and
16.  This is a much more extreme range than found in previous studies
with, for example, Baumgart \& Peterson (1986) finding only values in
the range 0.35--0.55.  The main discrepancy seems to lie in the very
elongated bars we find, with major axis FWHM, in two cases,  an
order of magnitude larger than minor axis FWHM ($b/a =$ 0.094 for
NGC 2529, 0.096 for UGC~6332).  In a few cases the minor axis FWHM is
barely larger than the seeing scale, which may lead to the minor axis values in
Table 3 being overestimated somewhat. Table 3 only shows minor axis
and major axis profiles for 24 galaxies. These are the only galaxies
for which we can get reliable values for these parameters. In the other barred
galaxies the bar is either similar in size to the seeing or easily
confused with the bulge. However, EA has still been calculated for all bars,
because this is unaffected by seeing effects as explained earlier.

Figure 17 shows the bar lengths in kpc plotted against the bulge
half-light radii.  It has been suggested, both on the basis of numerical
models (Sellwood 1981) and optical imaging (Athanassoula \& Martinet
1980, Baumgart \& Peterson 1986), that these quantities may be related,
with longer bars being found in galaxies with larger bulges.  However,
this correlation is not borne out by the present data.  We also find no
correlation between bar length and B/D ratio.

\section{CONCLUSIONS}

\noindent
We now summarise the main results from this study of spiral galaxy
bulges, disks and bars, concentrating on those which seem discrepant from
previously published results.

We find that bulge profiles are best fitted by $R^{1/2}$ or exponential
profiles, in agreement with de Jong (1996b).  However, we find no
correlation between the index of the best-fit bulge profile and 
Hubble type, B/D ratio or bulge luminosity.  B/D ratio also appears to
correlate only weakly with Hubble type, and we concur with de Jong
(1996b) in concluding that some other parameter must be the main
determinant of morphological type.

Approximately 1/3 of our sample show significant central reddening in
their J--K colours.  This central reddening can be extended, which we
take as evidence for a central starburst nucleus, or it can be
unresolved, which we interpret as a dust-embedded AGN.  The strongest
examples of these phenomena are IC~568, which appears to have a strong
nuclear starburst, and UGC~3707, which probably harbours a luminous
Seyfert nucleus.

Bars are found in 40 galaxies, out of a total sample size of 45.  Of the
barred galaxies, 9 were classified as unbarred in the RC3
on the basis of their optical morphologies.  Since our sample selection
is unbiassed with respect to bar strength, we believe this indicates that
most, and quite possibly all bright spirals are barred at some level.

We determine bar strengths using a newly-defined, distance- and
seeing-independent parameter called Equivalent Angle. This is preferred over 
the traditional bar-interbar method,  because  the latter
can  be  strongly  affected by seeing and other resolution  effects.
We find that bar
strength does not correlate with disk K-band surface brightness, but that
the strongest bars are found in a narrow, intermediate range of B/D.

We find two types of bar profiles, flat and exponential, in agreement with
the finding of Elmegreen et al. (1996).
However, we differ with Elmegreen et al. (1996) in finding
no correlation between bar type and Hubble type. The K-band minor-axis
bar profiles are significantly asymmetric, being generally steeper on the
leading edge than on the trailing edge, assuming that the arms have a
trailing geometry.  Again this asymmetry is not predicted by any of the
theoretical bar models, and was not found in the strongly barred galaxies
investigated by Ohta et al. (1990).  Finally, bar axial ratios were
found to extend over a much large range of values than claimed from
optical studies (Baumgart \& Peterson 1986), with major axes up to $\sim$10
times greater than minor axes.

\section{Acknowledgements}

We wish to thank Chris Collins and the referee for their useful comments 
in the preparation of this this paper.
The United Kingdom Infrared Telescope is operated by the Joint Astronomy
Centre on behalf of the U.K. Particle Physics and Astronomy Research
Council.
This research has made use of the NASA/IPAC Extragalactic Database (NED)
which is operated by the Jet Propulsion Laboratory, California Institute
of Technology, under contract with the National Aeronautics and Space
Administration.

\renewcommand{\baselinestretch}{1}

\clearpage
\begin{table*}
\caption{The effect of seeing on extracted bulge and disk properties of simulated galaxies. Tabulated are model input and fitted output values.}
\begin{center}
\footnotesize
\begin{tabular}{llllll}

Model 	& Bulge SB & Bulge radius  &  Bulge index & Disk CSB & Disk scale length \\
	& (counts) & (arcsec)	   &              & (counts) & (arcsec)		 \\
\hline
a	&25000/24865	&4.46/4.52	&0.5/0.494	&100/102	&8.58/8.69\\
b	&25000/26670	&1.49/1.37	&0.5/0.567	&100/105	&8.58/9.67\\
\hline
\end{tabular}
\normalsize
\end{center}
\end{table*}

\begin{table*}
\caption{Bulge and nuclear properties of the 45 spiral galaxies. The following data are included: Column 1, galaxy name; column 2, galaxy classification from RC3; column 3, galaxy redshift from NED; column 4, best-fit bulge index; column 5 bulge apparent magnitude at K; column 6 bulge half-light radius; and column 7, change in (J-K) colour within the central 5 arcsec in radius, which defines a central region large enough to be unaffected by variations in seeing.}
\begin{center}
\footnotesize
\begin{tabular}{lllllcl}
\hline
Galaxy & Hubble type & Redshift & Bulge   & Bulge     	&Bulge effective  	&Nuclear  \\
name   & (RC3)     & (kms$^{-1}$)& index   & app. mag.	&radius (arcsec)      &$\Delta(J-K)$\\
\hline
ESO 555-G013 & Sa	  &	2753   & --   & --    & --  & 0.1\\
IC 357	    & SBab	  &    	6261   & 0.50 & 11.18 & 2.17 & 0.25\\
IC 568	    & SBb	  &    	8720   & 0.74 & 12.40 & 1.74 & 1.30\\
IC 742	    & SBab	  &    	6425   & 0.50 & 12.20 & 2.09 & 0.35\\
IC 1196     & Sa	  &    	4885   & 0.49 & 12.54 & 2.00 & 0.30\\
IC 1764	    & SBb	  &    	5029   & 0.44 & 12.14 & 1.83 & 0.1\\
IC 1809	    & SBab	  &    	5578   & 0.37 & 11.37 & 1.74 & 0.3\\
IC 2363	    & SBbc	  &    	7616   & 0.53 & 12.45 & 1.89 & 0.2\\
IC 3692	    & SBa	  &    	6542   & 0.37 & 11.85 & 1.63 & 0.1\\
NGC 1219    & SAbc	  & 	6101   & 0.49 & 11.84 & 2.09 & 0.2\\
NGC 2416    & Scd	  & 	5101   & 0.50 & 10.96 & 1.72 & 0.1\\
NGC 2503    & SABbc	  & 	5506   & 0.50 & 13.32 & 1.32 & 0.3\\
NGC 2529    & SBd	  & 	5029   & 0.50 & 13.19 & 1.97 & 0.25\\
NGC 2628    & SABc	  & 	3622   & 0.60 & 12.45 & 1.72 & 0.15\\
NGC 3512    & SABc	  & 	1376   & 0.60 & 11.99 & 1.80 & --\\
NGC 5478    & SABbc	  & 	7539   & 0.90 & 11.83 & 2.80 & 0.15\\
NGC 5737    & SBb	  & 	9517   & 0.39 & 11.81 & 1.63 & 0.2\\
NGC 6347    & SBb	  & 	6144   & 0.50 & 13.52 & 2.06 & 0.1\\
NGC 6379    & Scd	  & 	5973   & 0.94 & 12.28 & 3.26 & 0.15\\
NGC 6574    & SABbc	  & 	2282   & 0.50 & 13.23 & 1.83 & 0.20\\
UGC 850	    & SAbc	  & 	17283   & 0.48 & 13.76 & 1.34 & 0.25\\
UGC 1478    & SBc	  & 	4846   & 0.90 & 12.47 & 5.41 & 0.0\\
UGC 1546    & SABc	  & 	2374   & 1.00 & 13.73 & 4.32 & 0.2\\
UGC 2303    & SABb	  & 	6404   & 0.50 & 12.21 & 1.26 & 0.05\\
UGC 2585    & SBb	  & 	6847   & 0.60 & 11.98 & 1.54 & 0.0\\
UGC 2705    & SBcd	  & 	6858   & 0.40 & 12.76 & 3.15 & 0.05\\
UGC 2862    & SABa	  & 	6644   & 0.54 & 10.53 & 2.77 & 0.05\\
UGC 3053    & Scd	  & 	2407   & 0.63 & 9.73  & 1.32 & --\\
UGC 3091    & SABd	  & 	5559   & 0.34 & 14.12 & 3.58 & 0.05\\
UGC 3171    & SBcd	  & 	4553   & 0.58 & 12.04 & 1.52 & 0.1\\
UGC 3233    & Scd	  & 	4662   & 0.95 & 13.63 & 4.92 & 0.05\\
UGC 3296    & Sab	  & 	4266   & 0.54 & 12.82 & 1.72 & 0.00\\
UGC 3578    & SBab	  & 	4531   & 0.43 & 11.44 & 1.57 & 0.15\\
UGC 3707    & Sab	  & 	5967   & 0.58 & 12.13 & 1.83 & 0.85\\
UGC 3806    & SBcd	  & 	5484   & 0.45 & 14.11 & 1.40 & 0.35\\
UGC 3839    & SBb	  & 	5267   & 0.49 & 12.65 & 2.00 & 0.30\\
UGC 3900    & SBb	  & 	8535   & 0.47 & 12.84 & 2.06 & 0.0\\
UGC 3936    & SBbc	  & 	4725   & 0.57 & 11.89 & 1.72 & 0.15\\
UGC 4643    & SAbc	  & 	7699   & 0.57 & 12.69 & 1.86 & 0.00\\
UGC 5434    & SABb	  & 	5580   & 0.43 & 13.42 & 2.15 & 0.00\\
UGC 6166    & Sbc	  & 	10182   & 0.57 & 15.21 & 2.89 & --\\
UGC 6332    & SBa	  & 	6245   & 0.52 & 10.74 & 1.74 & 0.05\\
UGC 6958    & SABbc	  & 	5931   & 0.97 & 12.49 & 2.43 & 0.05\\
UGC 8939    & SABb	  & 	7459   & 0.46 & 12.12 & 2.20 & 0.15\\
UGC 11524   & SAc	  & 	5257   & 0.87 & 12.28 & 4.66 & 0.15\\
\hline
\end{tabular}
\normalsize
\end{center}
\end{table*}

\begin{table*}
\caption{Disk and bar properties of the 45 spiral galaxies. The following data are presented: column 1, galaxy name; column 2, disk central surface brightness; column 3, disk scale length; column 4, disk apparent magnitude; column 5, bulge-to-disk ratio; column 6, bar strength; column 7, bar FWHM along major axis; column 8, bar FWHM along minor axis; and column 9, bar type.}
\begin{center}
\footnotesize
\begin{tabular}{lllllllll}
\hline
Galaxy Name & Disk CSB	& Disk scale & Disk  	& B/D	& Bar	& Bar long	& Bar short 	& Bar type\\ 
	    & (K mag) 	& length (arcsec)     & app. mag.    & 	& EA	& axis (arcsec)	& axis (arcsec)	&	\\
\hline
ESO 555 -G 013 & 18.20 & 9.7 & 10.69 & -- & -- & -- & -- & -- \\
IC 357	    & 16.47 & 8.6 & 10.02 & 0.34 & 23 & 13.3 & 2.2 & Flat \\
IC 568	    & 18.23 & 10.8 & 11.58 & 0.47 & 80 & 20.8 & 3.0 & Flat \\
IC 742	    & 18.88 & 24.7 & 10.46 & 0.2 & 18 & 15.7 & 2.6 & Flat \\
IC 1196     & 16.12 & 4.0 & 11.73 & 0.49 & 4 & -- & -- & -- \\
IC 1764	    & 18.12 & 12.0 & 10.78 & 0.29 & 27 & 11.2 & 3.6 & Flat \\
IC 1809	    & 16.90 & 6.4 & 11.27 & 0.91 & 16 & 5.9 & 3.1 & Exponential \\
IC 2363	    & 17.62 & 8.1 & 11.22 & 0.32 & 36 & 13.1 & 2.1 & Exponential \\
IC 3692	    & 19.48 & 7.5 & 10.92 & 0.42 & 42 & 8.3 & 1.25 & Exponential \\
NGC 1219    & 17.31 & 11.4 & 10.07 & 0.20 & 5 & -- & -- & -- \\
NGC 2416    & 17.50 & 12.3 & 10.51 & 0.66 & 9 & -- & -- & -- \\
NGC 2503    & 17.71 & 8.6 & 11.06 & 0.13 & 9 & 13.9 & 4.1 & Flat \\
NGC 2529    & 17.18 & 7.1 & 11.30 & 0.18 & 13 & 12.7 & 1.2 & Exponential \\
NGC 2628    & 17.34 & 11.4 & 10.10 & 0.12 & 12 & 6.6 & 1.7 & Flat \\
NGC 3512    & 16.56 & 11.4 & 9.40 & 0.09 & 12 & -- & -- & -- \\
NGC 5478    & 17.75 & 12.1 & 10.76 & 0.37 & 7 & -- & -- & -- \\
NGC 5737    & 17.31 & 11.5 & 10.59 & 0.33 & 63 & 10.1 & 2.3 & Exponential \\
NGC 6347    & 17.08 & 9.8 & 10.74 & 0.08 & 15 & -- & -- & -- \\
NGC 6379    & 17.83 & 14.3 & 10.26 & 0.16 & 7 & -- & -- & -- \\
NGC 6574    & 17.43 & 8.1 & 11.17 & 0.15 & 8 & -- & -- & -- \\
UGC 850	    & 18.36 & 8.6 & 11.98 & 0.19 & 6 & -- & -- & -- \\
UGC 1478    & 18.52 & 11.9 & 11.33 & 0.35 & 20 & 7.0 & 2.0 & Exponential \\
UGC 1546    & 18.52 & 8.5 & 11.89 & 0.18 & -- & -- & -- & -- \\
UGC 2303    & 17.48 & 7.6 & 11.07 & 0.35 & 34 & 6.1 & 1.0 & Exponential \\
UGC 2585    & 17.48 & 10.0 & 10.82 & 0.34 & 17 & 13.2 & 2.0 & Flat \\
UGC 2705    & 19.15 & 11.4 & 12.02 & 0.50 & 22 & 6.2 & 1.25 & Flat \\
UGC 2862    & 19.23 & 25.7 & 11.02 & 1.56 & 8 & -- & -- & -- \\
UGC 3053    & 16.35 & 12.1 & 9.27 & 0.65 & -- & -- & -- & -- \\
UGC 3091    & 18.91 & 9.7 & 11.84 & 0.15 & -- & -- & -- & -- \\
UGC 3171    & 17.57 & 8.9 & 11.06 & 0.40 & 8 & 4.8 & 1.1 & Flat \\
UGC 3233    & 19.45 & 11.4 & 12.43 & 0.33 & 10 & -- & -- & -- \\
UGC 3296    & 16.33 & 6.3 & 10.68 & 0.14 & 6 & -- & -- & -- \\
UGC 3578    & 16.02 & 7.8 & 10.31 & 0.35 & 39 & 15.5 & 3.0 & Flat \\
UGC 3707    & 18.02 & 18.0 & 11.03 & 0.36 & 62 & -- & -- & -- \\
UGC 3806    & 18.65 & 12.9 & 11.31 & 0.08 & 12 & 8.8 & 1.8 & Flat \\
UGC 3839    & 18.01 & 10.0 & 11.15 & 0.25 & 20 & 14.0 & 2.2 & Flat \\
UGC 3900    & 17.93 & 8.7 & 11.49 & 0.29 & 21 & 5.7 & 1.1 & Exponential \\
UGC 3936    & 16.79 & 8.6 & 10.34 & 0.24 & 16 & 8.4 & 1.6 & Exponential \\
UGC 4643    & 18.33 & 9.7 & 11.65 & 0.38 & 6 & -- & -- & -- \\
UGC 5434    & 18.16 & 15.7 & 10.61 & 0.08 & 5 & -- & -- & -- \\
UGC 6166    & 17.13 & 7.8 & 13.57 & 0.22 & -- & -- & -- & -- \\
UGC 6332    & 17.65 & 15.1 & 9.98 & 0.49 & 23 & 12.5 & 1.2 & Flat \\
UGC 6958    & 18.00 & 17.5 & 10.20 & 0.12 & 22 & 5.9 & 1.65 & Exponential \\
UGC 8939    & 18.65 & 14.2 & 11.09 & 0.39 & 20 & 7.8 & 1.4 & Exponential \\
UGC 11524   & 18.39 & 15.7 & 10.66 & 0.22 & 5 & -- & -- & -- \\
\hline
\end{tabular}
\normalsize
\end{center}
\end{table*}

\clearpage

\noindent
{\bf Figure 1.} Examples of the K-band galaxy light profiles produced by
the ellipse-fitting process. Lines show the best fits obtained using the
sum (dashed line) of an exponential disk (dotted line), and a generalised 
de Vaucouleurs profile (dash-dotted line) for the bulge.

\noindent
{\bf Figure 2.} Best fitting bulge index in the generalised de
Vaucouleurs profile, plotted against Hubble type.

\noindent
{\bf Figure 3.} Best fitting bulge index in the generalised de
Vaucouleurs profile, plotted against integrated K-band bulge absolute
magnitude.

\noindent
{\bf Figure 4.} Central J--K colour profiles in selected galaxies.

\noindent
{\bf Figure 5.} Disk K-band central surface brightness plotted against
disk absolute K magnitude.

\noindent
{\bf Figure 6.} Disk K-band central surface brightness plotted against
Hubble type.

\noindent
{\bf Figure 7.} Galaxy K-band bulge-to-disk ratio plotted against Hubble
type.

\noindent
{\bf Figure 8.} Disk exponential scale length plotted against bulge half-light
radius.

\noindent
{\bf Figure 9.} Disk K-band central surface brightness plotted against
disk exponential scale length.

\noindent
{\bf Figure 10.} Diagrammatic representation of equivalent angle.

\noindent
{\bf Figure 11.} The ranges of bar strengths, measured in averaged
equivalent angle (see text for definition). The black areas represent the
bar strengths of those galaxies with near neighbours.

\noindent
{\bf Figure 12.} Bar strength plotted against disk K-band central surface
brightness.

\noindent
{\bf Figure 13.} Bar strength plotted against K-band bulge-to-disk ratio.

\noindent
{\bf Figure 14.} Bar profiles without disk-- and 
bulge--subtraction (left column) and with disk-- and bulge--subtraction
for three representative galaxies.

\noindent
{\bf Figure 15.} Bar strength in equivalent angle plotted against radius for 
IC 568.

\noindent
{\bf Figure 16.} Four bar cross--sections with two typical cases (IC 1764 and 
IC 357) with steeper leading edges one case (IC 568) with a steeper trailing 
edge and one case (NGC 2529) approximately symmetric. The `leading' edge is 
plotted on the right.

\noindent
{\bf Figure 17.} Bar length plotted against bulge half--light radius.

\clearpage

\pagestyle{empty}

\begin{figure}
\includegraphics{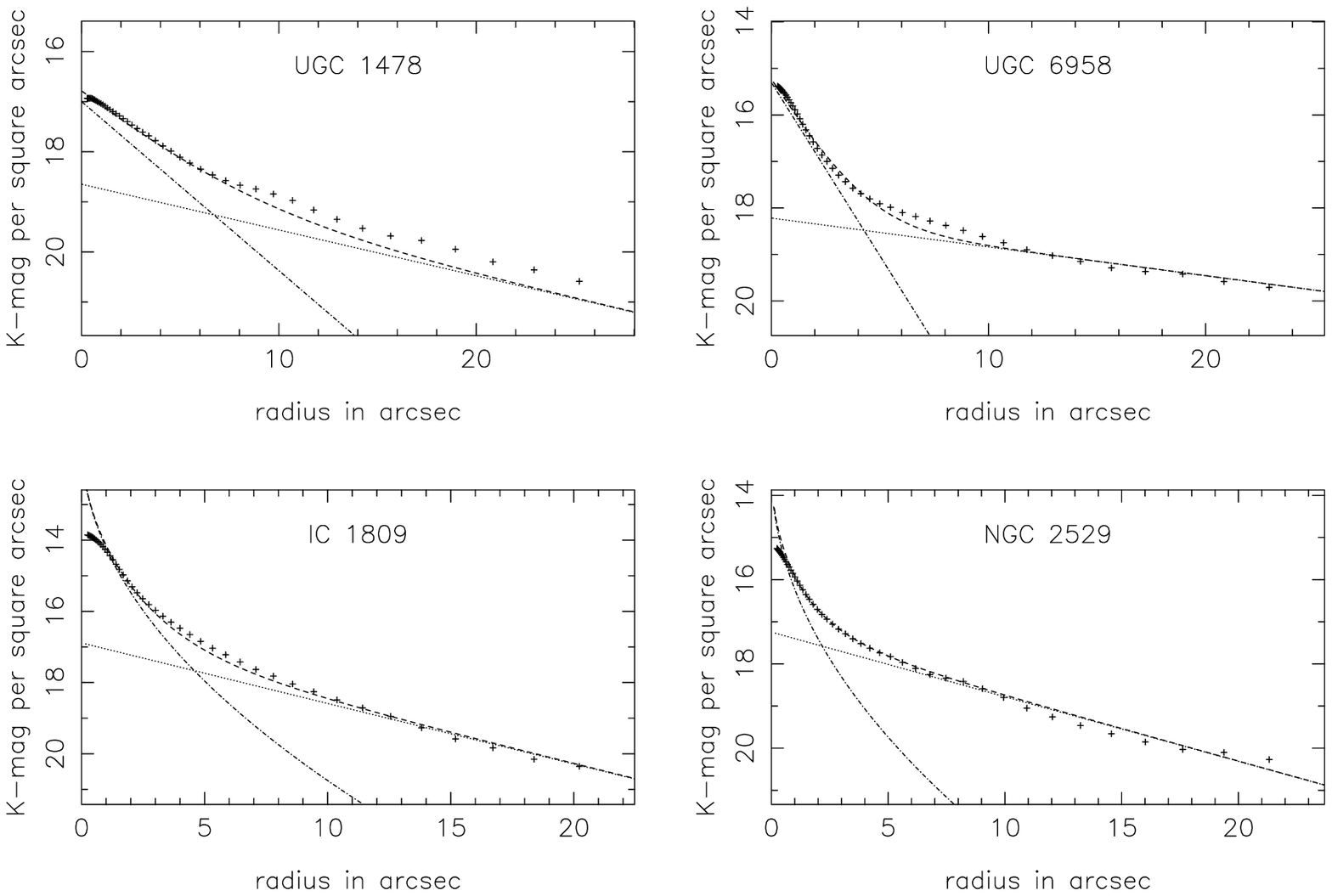}
\vspace*{20cm}
\end{figure}

\clearpage

\begin{figure}
\includegraphics{figure2.ps}
\vspace*{20.0cm}
\end{figure}

\clearpage

\begin{figure}
\includegraphics{figure3.ps}
\vspace*{15cm}
\end{figure}

\clearpage

\begin{figure}
\includegraphics{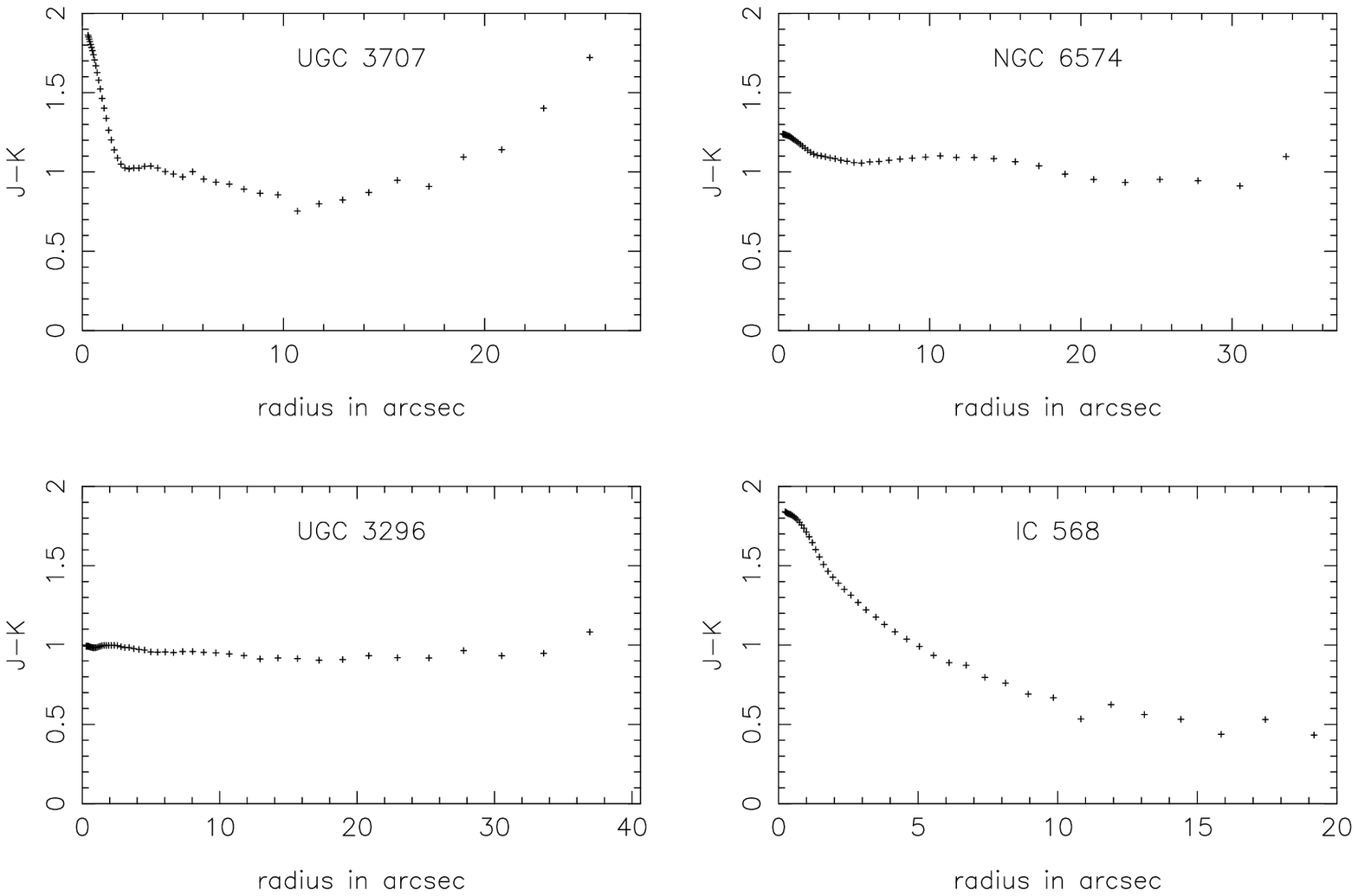}
\vspace*{15cm}
\end{figure}

\clearpage

\begin{figure}
\includegraphics{figure5.ps}
\vspace*{15cm}
\end{figure}

\clearpage

\begin{figure}
\includegraphics{figure6.ps}
\vspace*{15cm}
\end{figure}

\clearpage

\begin{figure}
\includegraphics{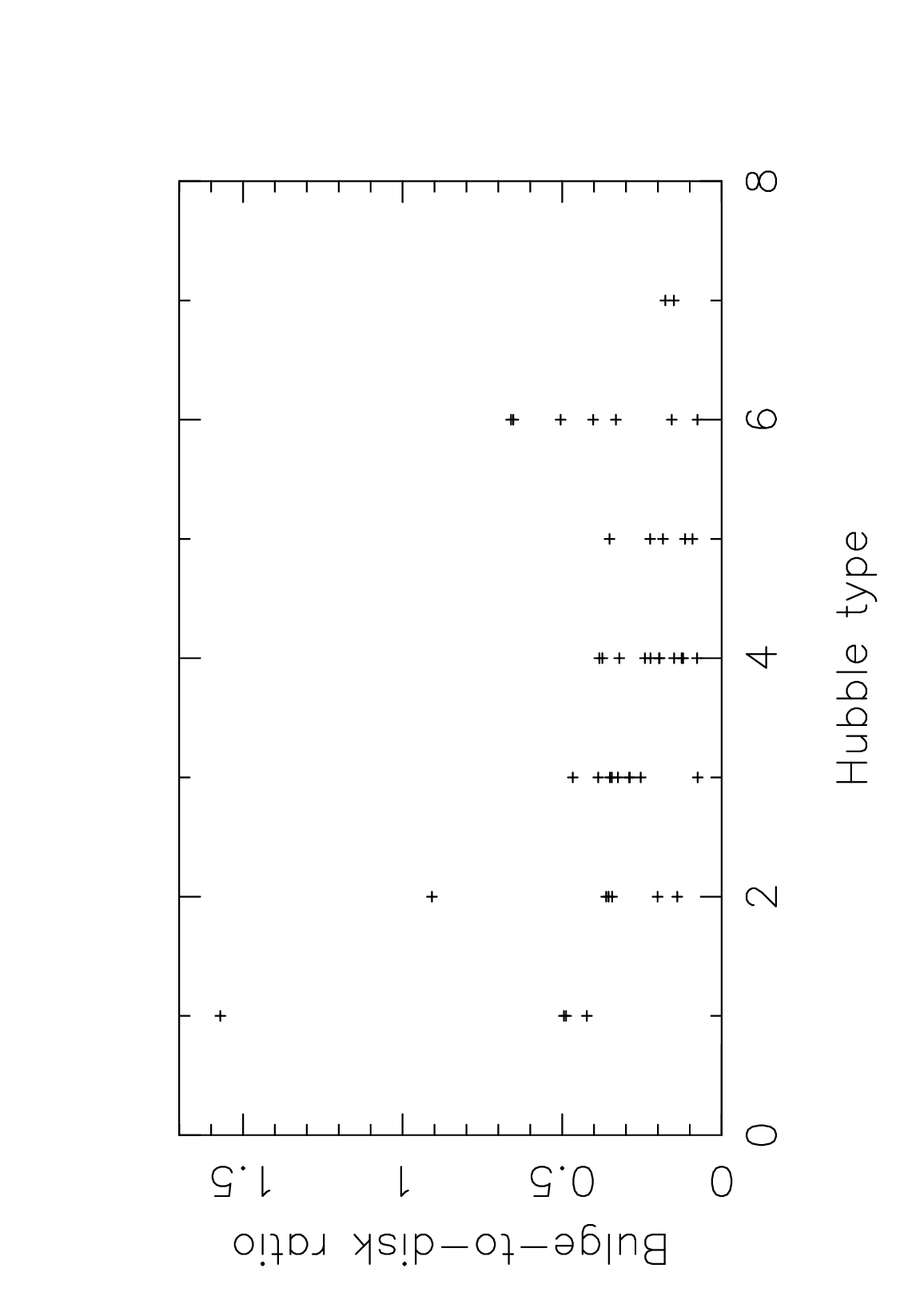}
\vspace*{15cm}
\end{figure}

\clearpage

\begin{figure}
\includegraphics{figure8.ps}
\vspace*{17cm}
\end{figure}

\clearpage

\begin{figure}
\includegraphics{figure9.ps}
\vspace*{15cm}
\end{figure}

\clearpage

\begin{figure}
\includegraphics{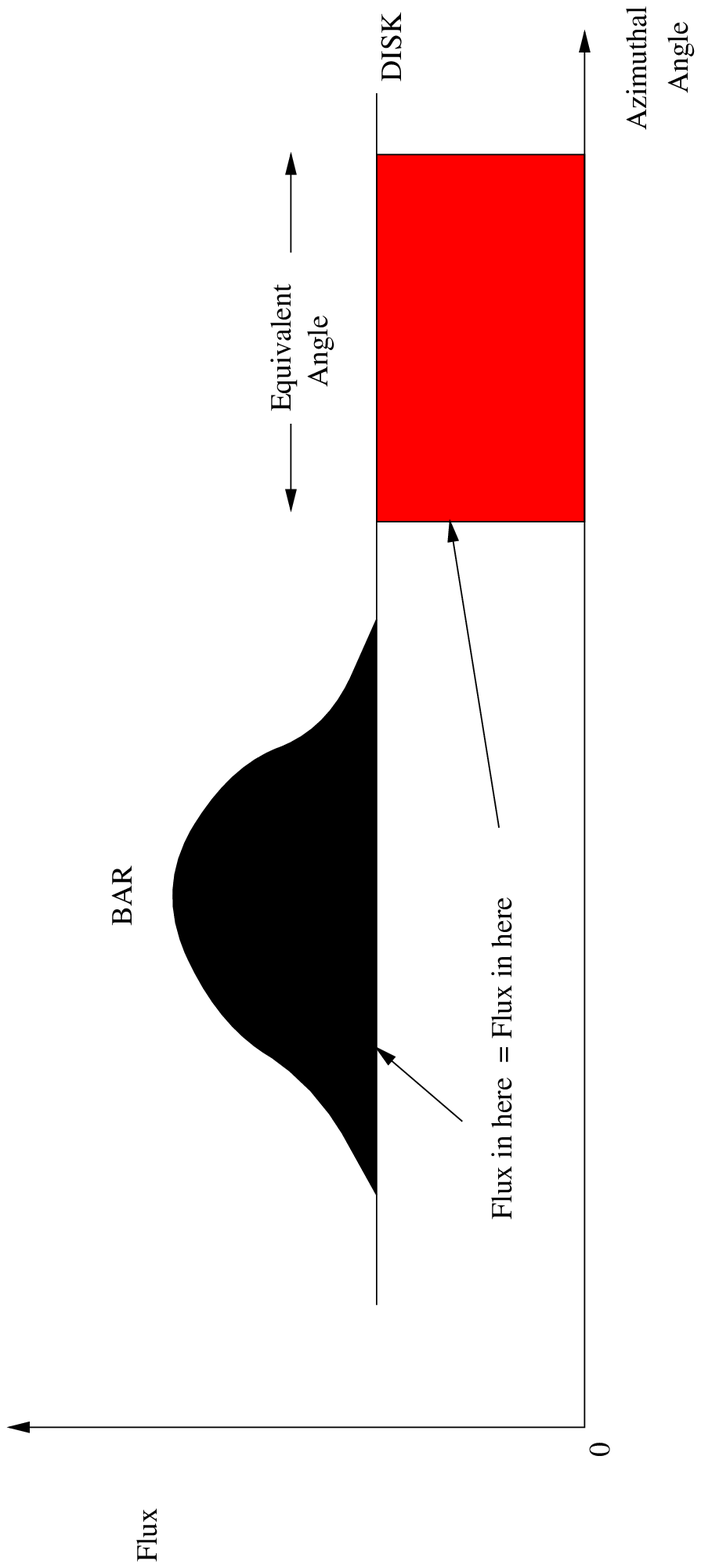}
\vspace*{15cm}
\end{figure}

\clearpage

\begin{figure}
\includegraphics{figure11.ps}
\vspace*{15cm}
\end{figure}

\clearpage

\begin{figure}
\includegraphics{figure12.ps}
\vspace*{15cm}
\end{figure}

\clearpage

\begin{figure}
\includegraphics{figure13.ps}
\vspace*{15cm}
\end{figure}

\clearpage

\begin{figure}
\includegraphics{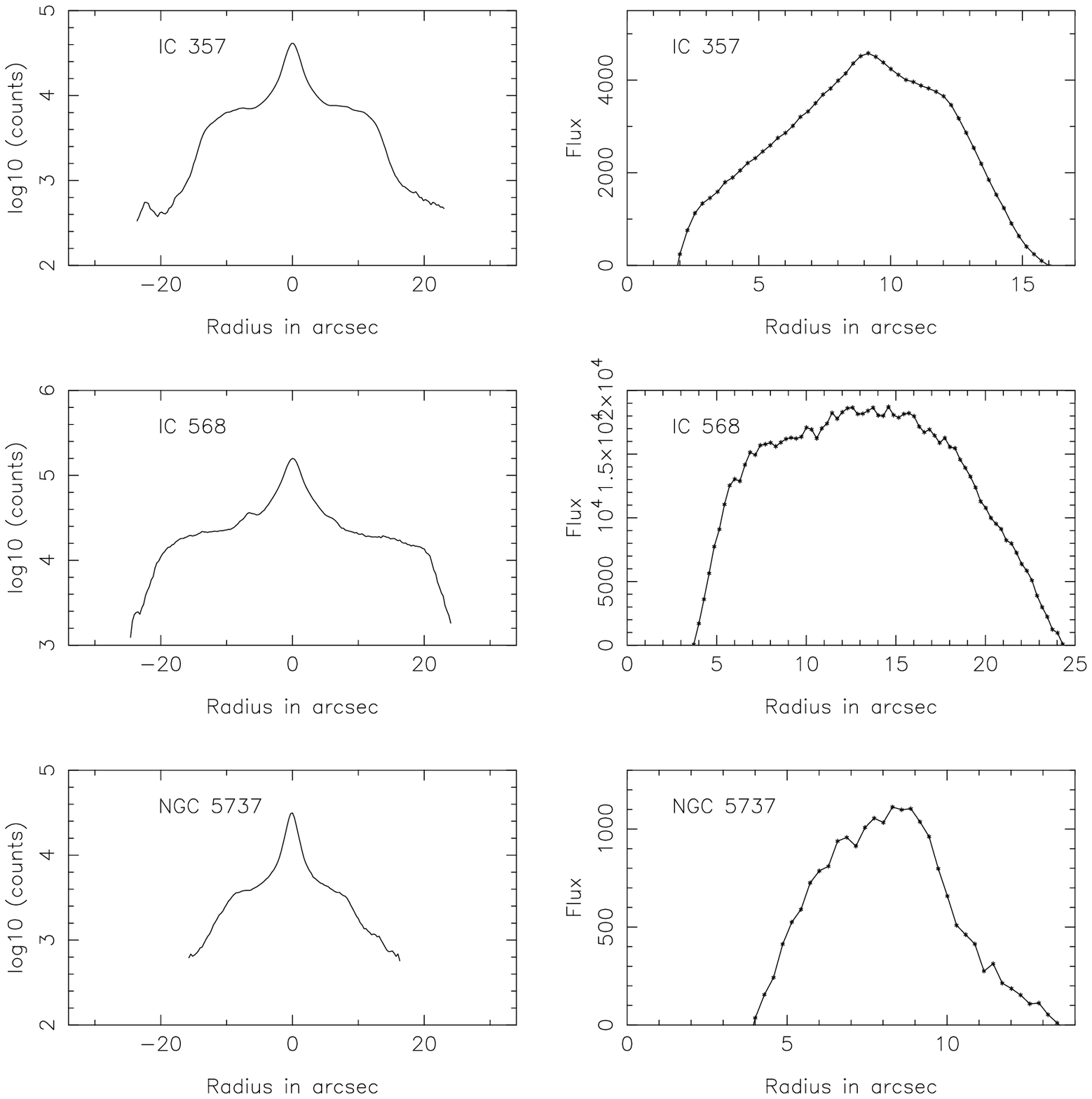}
\vspace*{15cm}
\end{figure}

\clearpage

\begin{figure}
\includegraphics{figure15.ps}
\vspace*{20cm}
\end{figure}

\clearpage

\begin{figure}
\includegraphics{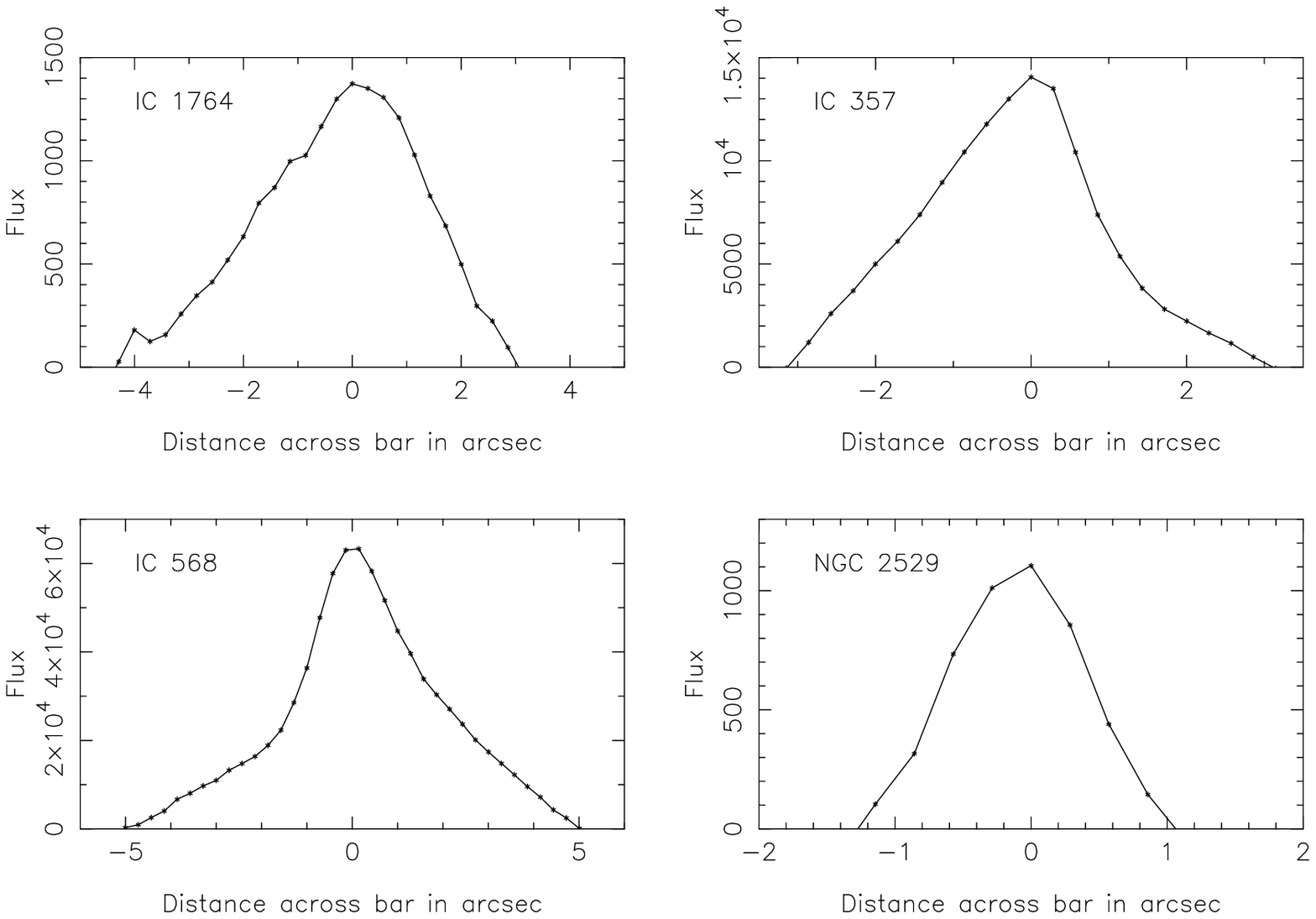}
\vspace*{20cm}
\end{figure}

\clearpage

\begin{figure}
\includegraphics{figure17.ps}
\vspace*{20cm}
\end{figure}

\clearpage

\end{document}